\documentstyle[fleqn,12pt]{article}

\newcommand{\mathrm}{\rm}

\newcommand{\nl}{\nonumber \\}

\newcommand{\bq}{\begin{equation}}
\newcommand{\eq}{\end{equation}}
\newcommand{\ba}{\begin{eqnarray}}
\newcommand{\ea}{\end{eqnarray}}

\begin{document}
\begin{flushright}
Freiburg--THEP 93/28\\
December 1993\\

to appear in the proceedings of the\\
Zvenigorod Workshop (1993)\\
\end{flushright}
\begin{center}
\vskip 1cm
{\large \bf The Hill Theorem}
\vskip 1cm
{\bf J. J. van der Bij }\\ \bigskip Albert Ludwigs Universit\"at Freiburg\\
Fakult\"at f\"ur Physik \\ H. Herder Strasse 3\\
79104 Freiburg i. Br.\\
\end{center}

\begin{abstract}
It is studied how strong interactions in the Higgs sector can lead to
deviations in vector boson selfcouplings. Normally such effects are
small due to Veltman's screening theorem. It is shown that strong
interactions are possible, if there is a hierarchy of strong interactions
in the Higgs sector. This is called Hill's theorem.

\end{abstract}

\section{Introduction \label{intr}}
The standard model of the weak interactions gives a satisfactory
description of all experiments up to date. The structure of the model is
also elegant in the sense that the interactions are determined by the gauge
principle and therefore of a geometric origin. The exception to this is
the Higgs sector of the theory. The Higgs sector is responsible for the large
number of free parameters, in casu the Yukawa couplings to the fermions
of the model. The Yukawa couplings give rise to the masses of the theory
via the mechanism of spontaneous symmetry breaking. There appears to exist
essentially no pattern to the structure of the fermion masses. It seems
therefore reasonable to ask whether the Higgs sector is fundamental.
Possibly strong interactions could be present. Since an explicit model
of such strong interactions is lacking, the best one can do as a start
is to study the theory without the Higgs explicitly present. If one removes the
Higgs the theory becomes nonrenormalizable and cut-off dependent
results appear. Such efffects can be estimated, for instance by taking
the Higgs particle to be very heavy and calculating the radiative corrections
to vectorboson properties [1]. Typically the generated effects are only
logarithmic
in the cut-off. Therefore no particularly strong interactions are present at
low energy. This is known as Veltman's theorem. Since the divergences can be
calculated directly in the non-renormalizable  model [2,3], this is at first
sight a universal feature. To study whether Veltman's theorem can be avoided
one therefore has to complicate the standard model.
A way to do this is to add a strongly interacting singlet sector to the theory.
The interactions in the Higgs sector can be affected by the presence of the
singlet sector. As a consequence the radiative corrections to vector boson
physics can be changed. Explicit calculations show that the resulting
interactions among the vector bosons can become strong. However for this
to happen a hierarchy of strong couplings has to be present in the theory.
This would appear to be a general feature, regardless of the exact nature
of the strong interactions. Therefore I claim that the following
holds (Hill's theorem) : "Veltman's screening theorem can be avoided if
there is a hierarchy of strong interactions in the Higgs sector".

\vskip 5mm

\section{Removing the Higgs particle}
The weak interactions are mediated by the exchange of massive vector bosons.
Within the standard model the mass of the vector bosons is due to the Higgs
mechanism. We are interested here in strongly interacting vector bosons
where no physical Higgs is present. Therefore we are interested in ways
to remove the Higgs particle from the theory. There are essentially two
approaches possible. One is the traditional approach making the Higgs
very heavy. Since the mass of the Higgs depends on the coupling making the
Higgs heavy corresponds to a srongly coupled theory. The standard model
is a gauged linear $\sigma$ model with the Lagrangian:
\ba
{\cal L} = -\frac{1}{2} (D_{\mu} \Phi)^{\dagger}(D^{\mu} \Phi)
-\frac {\lambda}{8}(\Phi^{\dagger} \Phi - f^2)^2
\ea

\ba
\Phi = (\sigma + i\vec \tau \cdot \vec \phi)
\left(\begin{array}{c} 1 \\ 0 \end{array}\right)
\ea

At the tree level, the Higgs particle can be removed
by taking the limit $\lambda \rightarrow \infty$ or,
equivalently, $m_H^2 \rightarrow \infty$. The standard model
then turns into a gauged nonlinear $\sigma$-model

\ba
{\cal L} =\frac {f^2}{4} Tr(D_{\mu} U)^{\dagger}(D^{\mu} U)
\ea

\ba
U = \sqrt {1-\vec \pi^2} + i \vec \tau \cdot \vec \pi
\ea

\ba \vec \pi = \frac {\vec \phi}{f}
\ea

which is equivalent to massive Yang-Mills theory. It can be seen
on the formal level that the standard model reduces to (3)
in the limit $\lambda \rightarrow \infty$ by noticing that the
potential acts like a constraint in this case.\nl
A second \cite{bij}, more speculative way to remove the Higgs boson relies
on the possibility of having a large nonminimal coupling
$\xi \Phi^{\dagger} \Phi R$ of the Higgs
boson to gravity. As a consequence there is a large wave function
renormalization of the Higgs boson by a factor $1/\sqrt{1+12\xi}$.
This reduces the coupling of the Higgs boson to gravitational strength,
so that the Higgs effectively disappears from the theory.\nl
Given the removal of the Higgs boson from the theory there are two ways
to perform the radiative corrections. One is to calculate directly
in the non-linear model [2,3]. This calculation gives logarithmic divergences
that show up as poles in $1/(n-4)$ in dimensional regularization. The other
way is to calculate those physical effects in the linear model, that grow
with the Higgs mass [1]. Only a logarithmic growth $log(m_H)$ is present
in experimentally measurable quantities. These logaritmic effects are in
one to one correspondence with the divergences in the non-linear model
with the replacement $log(m_H) = 1/(n-4)$.
If one ignores the hypercharge field the effects can be summarized by the
following effective Lagrangian
\ba
{\cal L}_{eff} = \alpha_1 Tr(V_{\mu} V^{\mu})Tr(V_{\nu} V^{\nu})
+ \nl \alpha_2 Tr(V_{\mu} V^{\nu})Tr(V_{\mu} V^{\nu})
+ g \alpha_3 Tr( F_{\mu \nu} [V^{\mu},V^{\nu}])
\ea
where
\ba
V_{\mu} = (D_{\mu}U)U^{\dagger}
\ea
and
\ba
F_{\mu \nu} = (\partial_{\mu} - \frac{ig}{2} \vec W_{\mu} \cdot \vec \tau)
\frac {\vec W_{\nu} \cdot \vec \tau}{2i} -
(\mu \leftrightarrow \nu)
\ea

Explicit calculation in the linear model gives
\ba
\alpha_{1} =  \frac {1}{384 \pi^2} \ln (m_H^2/M^2_W) +{\cal O}(1)
\ea
\ba
\alpha_{2} =  \frac {1}{192 \pi^2} \ln (m_H^2/M^2_W) +{\cal O}(1)
\ea
\ba
\alpha_{3} = - \frac {1}{384 \pi^2} \ln (m_H^2/M^2_W) +{\cal O}(1)
\ea
The fact that these corrections are small, in particular that no correction
growing like $m_H^2$  is present is called Veltman's screening theorem.
In order to avoid this, one has to make changes in the Higgs sector.

\vskip 5mm

\section{Hill's theorem}
In order to avoid the consequences of the screening theorem,  changes
to the theory are necessary. The absence of quadratic divergences is
due to a delicate cancellation between radiative corrections and counterterms
in the Higgs propagator. These cancellations can be removed by adding extra
interactions in the Higgs sector. The simplest way to do this is to add
a strongly interacting singlet to the model [5]. The Lagrangian is given by

\ba
{\cal L} = -\frac{1}{2}(D_{\mu} \Phi)^{\dagger}(D^{\mu} \Phi)
-\frac {1}{2}(\partial_{\mu} x)^2 - \frac {\lambda_1}{8}
(\Phi^{\dagger} \Phi -f_1^2)^2 \nl -
\frac {\lambda_2}{8}(2f_2 x-\Phi^{\dagger}\Phi)^2 + {\cal L}_{gauge}
\ea
Within this model we are interested in the limit
 $\lambda_2 >> \lambda_1 >> 0$. In this limit the effects of the extra field
can in principle become strong, so that radiative effects can also feed down
to the interactions in the vector boson sector. The result of the explicit
calculation of the one loop radiative corrections is very simple
if one takes the limit $f_2 >> f_1$ at the end of the calculation.
In this limit the whole effect of the interactions beyond the standard model
can be summarized by the parameter $\beta$:
\ba
\beta = 128 \pi^2 ( \alpha_2 - 2\alpha_1)= \lambda_2/\lambda_1
\ea
The parameter $\beta$ thus takes arbitrary values in this
model. The same happens in an alternative model [6]. However in both cases
large effects only appear when there is a hierarchy of strong interactions.
This appears therefore to be a general feature.
As a consequence one can claim  the following  (Hill's theorem):
"Strong interactions can appear in the vector boson sector when there is a
hierarchy of interaction strengths in the Higgs sector".
One could object that the
model looks unnatural. However one presumably should not consider the x field
to be fundamental, but only as an effective description for underlying strong
dynamics. In that case the theorem could play a role in technicolor dynamics.
The parameter $\beta$ is known from pion physics [7,8]. It is the parameter
that is
responsible for the formation of I=1 bound states. The consequence of strong
interactions would be the formation of a I=1 bound state of vector bosons.
In principle such a state could be seen at the Tevatron if it is light
enough. Present limit are very weak, $\beta <500$ [9]. As a rough estimate this
corresponds to strong technicolor dynamics at a scale of about 20-50 TeV .

\vskip 5mm
\newpage

\end{document}